# Comments on Model-free temperature scaling for heat capacity, Drebushchak V. A. Journal of Thermal Analysis and Calorimetry, (2017) 130(1), 5–13

*I.H. Umirzakov*

Institute of Thermophysics, Pr. Lavreneva St., 1, Novosibirsk, Russia, 630090

**Abstract** It is shown that the isobaric heat capacity of chalcogenides $LiInS_2$, $LiInSe_2$, $LiGaS_2$, $LiGaSe_2$ and $LiGaTe_2$ can be described by the Debye and Einstein models for the phonon frequency spectrum within their uncertainties; the models give the results for the isochoric heat capacity which are close to each other; the models give the close results for the difference between the isobaric and isochoric heat capacities; the isobaric heat capacities of the isostructural $LiInS_2$, $LiInSe_2$, $LiGaS_2$ and $LiGaSe_2$ as the functions of the temperature reduced to the Debye (Einstein) temperature are described by single Debay (Einstein) equation for the isobaric heat capacity; the isochoric heat capacities of $LiInS_2$, $LiInSe_2$, $LiGaS_2$, $LiGaSe_2$ and $LiGaTe_2$ (which has another structure than $LiInS_2$, $LiInSe_2$, $LiGaS_2$ and $LiGaSe_2$ [1]) as the functions of the temperature reduced to the Debye (Einstein) temperature are described by the Debye (Einstein) equation for the isochoric heat capacity. It is shown also that the Debye and Einstein equations for the isochoric heat capacity of $LiInS_2$, $LiInSe_2$, $LiGaS_2$, $LiGaSe_2$ and $LiGaTe_2$ give the same results if the means of the squares of the frequencies of the Debye and Einstein spectra are equal to each other, and the Debye and Einstein equations for the isobaric heat capacity of $LiInS_2$, $LiInSe_2$, $LiGaS_2$ and $LiGaSe_2$ as the functions of the temperature reduced to the Debye or Einstein temperature give the same results.

**Introduction**

Recently the isobaric heat capacity $C_P(T)$ of five chalcogenides $LiInS_2$, $LiInSe_2$, $LiGaS_2$, $LiGaSe_2$ and $LiGaTe_2$ measured with DSC in a temperature range from 180 to 460 K was reported [1]. The model-free temperature scaling for the isobaric heat capacity of these substances using the data [1] were presented in [2]. The substances are close in their compositions and four of them are isostructural [1,2]. According to [2] the temperature dependence of their isobaric heat capacities cannot be treated with any existent heat capacity model. The main reason of this problem is the excess heat capacity as compared with $3R$ per mole of atoms, the upper theoretical limit for vibrational heat capacity of solids. Most experimental results after DSC measurements of $I-III-VI_2$ compounds turned out to show the excess of $C_P(T)$ over $3R$ per mole of atoms ($12R$ per mole for $I-III-VI_2$ compounds with four atoms) [1-8]. $LiGaTe_2$, $LiInSe_2$ and $LiGaSe_2$ have the heat capacities at 460 K exceeding $12R$, $LiInS_2$ has its heat capacity equal to the upper limit, and one $LiGaS_2$ is about 1% less than the upper limit [1,2]. Similar DSC results with the heat capacity exceeding the $3R$ limit were published for many similar chalcogenides: $AgGaSe_2$ and $AgInS_2$ [3]; $LiInS_2$, $LiInSe_2$, and $LiInTe_2$ [4]; $ZnSnAs_2$ [5]; $CuGaS_2$, $CuGaTe_2$, $CuInS_2$, $CuInSe_2$, and $CuInTe_2$ [6]; and $ArGaS_2$ [7]. Adiabatic calorimetry for $ArInTe_2$ did show the exceeding of the $3R$ limit even below 300 K [8].

We derive explicitly the equations for the isobaric heat capacity for the Debye and Einstein models of the phonon frequency spectrum in the present paper. We show that: the data [1] for five chalcogenides $LiInS_2$, $LiInSe_2$, $LiGaS_2$, $LiGaSe_2$ and $LiGaTe_2$ are described by the Debye and Einstein models for the phonon frequency spectrum within their uncertainties; the models give the results for the isochoric heat capacity which are close to each other; the models give the close results for the difference between the isobaric and isochoric heat capacities; the isobaric heat capacities of the isostructural $LiInS_2$, $LiInSe_2$, $LiGaS_2$ and $LiGaSe_2$ as the functions of the temperature reduced to the Debye (Einstein) temperature are described by single Debye (Einstein) equation for the isobaric heat capacity; the isochoric heat capacities of $LiInS_2$, $LiInSe_2$, $LiGaS_2$, $LiGaSe_2$ and $LiGaTe_2$ (which has another structure than $LiInS_2$, $LiInSe_2$, $LiGaS_2$ and $LiGaSe_2$ [1]) as the functions of the temperature reduced to the Debye (Einstein) temperature are described by the Debye (Einstein) equation for the isochoric heat capacity. It is shown also that the Debye and Einstein equations for the isochoric heat capacity of $LiInS_2$, $LiInSe_2$, $LiGaS_2$, $LiGaSe_2$ and $LiGaTe_2$ give the same results if the means of the squares of the frequencies of the Debye and Einstein spectra are equal to each other, and the Debye and Einstein equations for the isobaric heat capacity of $LiInS_2$, $LiInSe_2$, $LiGaS_2$ and $LiGaSe_2$ as the functions of the temperature reduced to the Debye or Einstein temperature give the same results.

**The predictions of the Debye and Einstein models for the isobaric heat capacity**

We consider the harmonic crystal consisting of $N$ atoms and having a volume $V$ with frequency spectrum $g(v,\omega)$, where $v = V/N$ is the volume per atom and $\omega$ is the frequency. The Debye spectrum $g_D(v,\omega)$ is defined by [9-11]

$$g_D(v,\omega) = 9\omega^2 \cdot \theta[\omega_D(v) - \omega]/\omega_D^3(v), \qquad (1)$$

where $\omega_D(v)$ is the Debye frequency, $\theta(x)$ is the Heaviside step function: $\theta(x) = 0$ if $x \leq 0$ and $\theta(x) = 1$ if $x > 0$. The isobaric $C_{VD}(v,T)$ and isochoric $C_{PD}(v,T)$ heat capacities per atom and the difference between them $\Delta C_D(v,T) = C_{PD}(v,T) - C_{VD}(v,T)$ are defined by (see Eqs. 12, 25 and 26 in **Appendix**)

$$C_{VD}(v,T) = 9k\left(\frac{T}{\theta_D}\right)^3 \cdot \int_0^{\theta_D/T} \frac{x^4 e^x}{(e^x - 1)^2} dx, \qquad (2)$$

$$C_{PD}(v,T) = C_{VD}(v,T) + \frac{(T/\theta_D) \cdot [C_{VD}^2(v,T)/k]}{a_D + \frac{9b_D}{8} + \frac{9b_D}{4}\left[\exp\left(\frac{\theta_D}{T}\right) - 1\right]^{-1} - \frac{T}{\theta_D}\left(1 - \frac{b_D}{4}\right) \cdot \frac{C_{VD}(v,T)}{k}}, \qquad (3)$$

$$\Delta C_D(v,T) = \frac{(T/\theta_D) \cdot [C_{VD}^2(v,T)/k]}{a_D + \frac{9b_D}{8} + \frac{9b_D}{4}\left[\exp\left(\frac{\theta_D}{T}\right) - 1\right]^{-1} - \frac{T}{\theta_D}\left(1 - \frac{b_D}{4}\right) \cdot \frac{C_{VD}(v,T)}{k}}, \qquad (4)$$

where $\theta_D = \theta_D(v) = \hbar\omega_D(v)/k$ is the Debye temperature, $\theta'_D = \theta'_D(v) = \hbar\omega'_D(v)/k$, $\theta''_D = \hbar\omega''_D(v)/k$, $u'_0 = u'_0(v) = du_0/dv$, $u''_0 = u''_0(v) = d^2u_0(v)/dv^2$, $\omega'_D = \omega'_D(v) = d\omega_D(v)/dv$, $\omega''_D = \omega''_D(v) = d^2\omega_D(v)/dv^2$, $a_D = a_D(v) = \omega_D u''_0/\hbar(\omega'_D)^2$, $b_D = b_D(v) = \omega_D \omega''_D/(\omega'_D)^2$, $T$ is the temperature, $u_0(v)$ is the internal energy per atom at $T=0$, $\hbar$ is the Planck constant and $k$ is Boltzmann constant.

The Einstein frequency spectrum is defined by [10,12]

$$g_E(v,\omega) = 3\delta[\omega - \omega_E(v)], \tag{5}$$

where $\omega_E(v)$ is the Einstein frequency. The isobaric $C_{VE}(v,T)$ and isochoric $C_{PE}(v,T)$ heat capacities per atom and the difference between them $\Delta C_E(v,T) = C_{PE}(v,T) - C_{VE}(v,T)$ are defined by (see Eqs. 33 and 38 in **Appendix**)

$$C_{VE}(v,T) = 3k\left(\frac{\theta_E(v)}{T}\right)^2 \exp\left(\frac{\theta_E(v)}{T}\right)\left[\exp\left(\frac{\theta_E(v)}{T}\right) - 1\right]^{-2}, \tag{6}$$

$$C_{PE}(v,T) = C_{VE}(v,T) + \frac{[T/\theta_E(v)] \cdot [C_{VE}^2(v,T)/k]}{a_E(v) + \frac{3b_E(v)}{2} + 3b_E(v)\left[\exp\left(\frac{\theta_E(v)}{T}\right) - 1\right]^{-1} - \frac{T}{\theta_E(v)} \cdot \frac{C_{VE}(v,T)}{k}}, \tag{7}$$

$$\Delta C_E(v,T) = \frac{[T/\theta_E(v)] \cdot [C_{VE}^2(v,T)/k]}{a_E(v) + \frac{3b_E(v)}{2} + 3b_E(v)\left[\exp\left(\frac{\theta_E(v)}{T}\right) - 1\right]^{-1} - \frac{T}{\theta_E(v)} \cdot \frac{C_{VE}(v,T)}{k}}, \tag{8}$$

where $\theta_E = \theta_E(v) = \hbar\omega_E(v)/k$ is the Einstein temperature, $\theta'_E = \hbar\omega'_E(v)/k$, $\theta''_E = \hbar\omega''_E(v)/k$, $\omega'_E = \omega'_E(v) = d\omega_E(v)/dv$, $\omega''_E = \omega''_E(v) = d^2\omega_E(v)/dv^2$, $a_E = a_E(v) = \omega_E u''_0/\hbar(\omega'_E)^2$ and $b_E = b_E(v) = \omega_E \omega''_E/(\omega'_E)^2$.

**Comparison with experimental data**

The products of $4N_A$ ($N_A$ is the Avogadro number) and the heat capacities were used in order to compare the predictions of the Debye and Einstein models with the data [1] for chalcogenides consisting of four atoms.

The isobaric $\overline{C}_{PD}(p,T)$ and isochoric $\overline{C}_{VD}(p,T)$ heat capacities and the difference between them $\Delta\overline{C}_D(p,T)$ as the functions of the independent variables $p$ and $T$ for the Debye model are defined by

$$\overline{C}_{PD}(p,T) \equiv C_{PD}(v,T)\big|_{v=v_D(p,T)}, \; \overline{C}_{VD}(p,T) \equiv C_{VD}(v,T)\big|_{v=v_D(p,T)}, \; \Delta\overline{C}_D(p,T) \equiv \Delta C_D(v,T)\big|_{v=v_D(p,T)}, \tag{9}$$

where $v_D(p,T)$ is defined from following equation (see Eq. 13 in **Appendix**)

$$p = \left[-u'_0(v) - \frac{9}{8}\hbar\omega'_D(v) - \frac{9\hbar\omega'_D(v)}{\omega_D^4(v)}\int_0^{\omega_D(v)} \omega^3\left[\exp\left(\frac{\hbar\omega}{kT}\right) - 1\right]^{-1} d\omega\right]_{v=v_D(p,T)}. \tag{10}$$

The isobaric $\overline{C}_{PE}(p,T)$ and isochoric $\overline{C}_{VE}(p,T)$ heat capacity per atom and the difference between them $\Delta \overline{C}_E(p,T)$ as the functions of the independent variables $p$ and $T$ for the Einstein model are defined by

$$\overline{C}_{PE}(p,T) \equiv C_{PE}(v,T)\big|_{v=v_E(p,T)}, \quad \overline{C}_{VE}(p,T) \equiv C_{VE}(v,T)\big|_{v=v_E(p,T)}, \quad \Delta \overline{C}_E(p,T) \equiv \Delta C_E(v,T)\big|_{v=v_E(p,T)}, \quad (11)$$

where $v_E(p,T)$ is defined from following equation (see Eq. 34 in **Appendix**)

$$p = \left[ -u'_0(v) - \frac{3\hbar \omega'_E(v)}{2} - 3\hbar \omega'_E(v) \left[ \exp\left(\frac{\hbar \omega_E(v)}{kT}\right) - 1 \right]^{-1} \right]\Bigg|_{v=v_E(p,T)}. \quad (12)$$

In order to describe the data [1] we assume that $\theta_D$, $a_D$, $b_D$, $\theta_E$, $a_E$ and $b_E$ are approximately constant in the interval $180\,K \leq T \leq 460\,K$ at $p = const$. This assumption is usually used for description of the isobaric heat capacity of the solids [9-11]. The best fit values of the parameters $\theta_D$, $a_D$, $b_D$, $\delta_D$ $\theta_E$, $a_E$, $b_E$ and $\delta_E$, where

$$\delta_{D,E} = \left[ \frac{1}{M-1} \sum_{i=1}^{M} \left( \frac{C_{PD,PE}(v,T_i)}{C_P(v,T_i)} - 1 \right)^2 \right]^{1/2} \cdot 100\%, \quad (13)$$

$M = 16$ is the number experimental points [1], are presented in Table 1.

**Table 1.** The values of the parameters $\theta_D$, $a_D$, $b_D$, $\theta_E$, $a_E$ and $b_E$ of five chalcogenides.

| Compound | $\theta_D$, K | $a_D$ | $b_D$ | $\delta_D$, % | $\theta_E$, K | $a_E$ | $b_E$ | $\delta_E$, % |
|---|---|---|---|---|---|---|---|---|
| LiInS2 | 443.246 | 62.391 | 2.687 | 0.25 | 338.545 | 88.334 | 1.708 | 0.32 |
| LiInSe2 | 344.161 | 69.306 | 3.265 | 0.20 | 263.996 | 97.034 | 1.789 | 0.23 |
| LiGaS2 | 478.759 | 64.101 | 2.366 | 0.26 | 364.577 | 94.050 | 1.376 | 0.36 |
| LiGaSe2 | 372.579 | 67.480 | 2.702 | 0.30 | 285.722 | 93.553 | 1.468 | 0.34 |
| LiGaTe2 | 246.143 | 53.683 | 2.726 | 0.17 | 195.384 | 57.670 | 3.462 | 0.17 |

The comparison of the Debye and Einstein theoretical isobaric heat capacities $C_{PD}(v,T)$ and $C_{PE}(v,T)$, which are calculated using Eqs. 3 and 7 and the values of the parameters $\theta_D$, $a_D$, $b_D$, $\theta_E$, $a_E$ and $b_E$ from Table 1, with the experimental data on $C_P(v,T)$ [1] for chalcogenides $LiInS_2$, $LiInSe_2$, $LiGaS_2$, $LiGaSe_2$ and $LiGaTe_2$ in the interval $180\,K \leq T \leq 460\,K$ are presented on Fig. 1.

The relative deviations $\Delta_D = C_{PD}(v,T)/C_P(v,T) - 1$ and $\Delta_E = C_{PE}(v,T)/C_P(v,T) - 1$ of the isochoric heat capacities $C_{VD}(v,T)$ and $C_{VE}(v,T)$, which are calculated using Eqs. 3 and 7 and the values of the parameters $\theta_D$, $a_D$, $b_D$, $\theta_E$, $a_E$ and $b_E$ from Table 1, from the data on $C_P(v,T)$ [1] for $LiInS_2$, $LiInSe_2$, $LiGaS_2$, $LiGaSe_2$ and $LiGaTe_2$ are presented on Fig. 2.

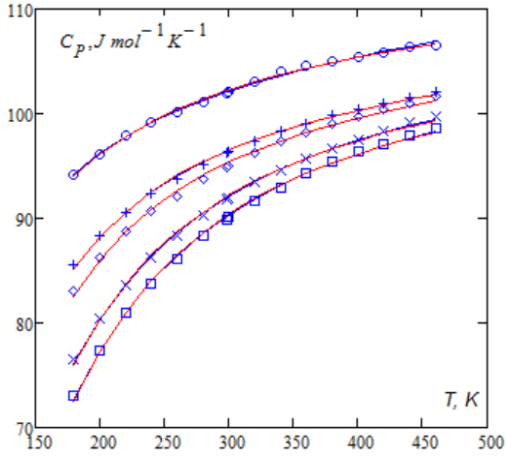

Fig. 1. The comparison of the Debay and Einstein theoretical isobaric heat capacities $C_{PD}(v,T)$ (the solid blue lines) and $C_{PE}(v,T)$ (the solid red lines) (Eqs. 3 and 7) with the experimental data on $C_P(v,T)$ [1] for $LiInS_2$ (the crosses), $LiInSe_2$ (the pluses), $LiGaS_2$ (the squares), $LiGaSe_2$ (the diamonds) and $LiGaTe_2$ (circles).

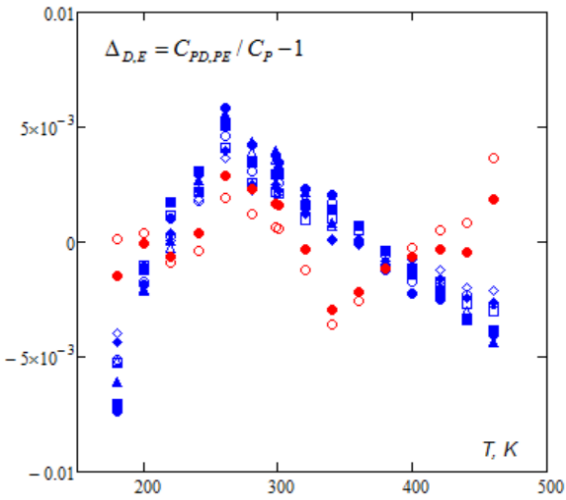

Fig. 2. The relative deviations $\Delta_D = C_{PD}/C_P - 1$ (the open symbols) and $\Delta_E = C_{PE}/C_P - 1$ (the filled symbols) of the isochoric heat capacities $C_{VD}$ and $C_{VE}$ (the solid red lines) (Eqs. 3 and 7) from the experimental data on $C_P(v,T)$ [1] for $LiInS_2$ (the blue squares), $LiInSe_2$ (the blue diamonds), $LiGaS_2$ (the blue circles), $LiGaSe_2$ (the blue triangles) and $LiGaTe_2$ (the red circles).

As one can see from Eq. 13, Table 1 and Figs. 1 and 2 the Debye and Einstein models describe the experimental data [1] within of their experimental uncertainties which are about 1%.

Fig. 3 presents the temperature dependence of the isochoric heat capacities $C_{VD}(v,T)$ and $C_{VE}(v,T)$ for $LiInS_2$, $LiInSe_2$, $LiGaS_2$, $LiGaSe_2$ and $LiGaTe_2$ which are calculated from Eqs. 2 and 6 using the values of the parameters $\theta_D$ and $\theta_E$ from Table 1.

The temperature dependence of the differences $\Delta C_D(v,T)$ and $\Delta C_E(v,T)$ between isobaric and isochoric heat capacities for $LiInS_2$, $LiInSe_2$, $LiGaS_2$, $LiGaSe_2$ and $LiGaTe_2$, which are calculated from Eqs. 4 and 8 using the values of the parameters $\theta_D$, $a_D$, $b_D$, $\theta_E$, $a_E$ and $b_E$ from Table 1, are presented on Fig. 4. One can see from Fig. 4 that the differences increase with increasing temperature in the interval $180\,K \leq T \leq 460\,K$.

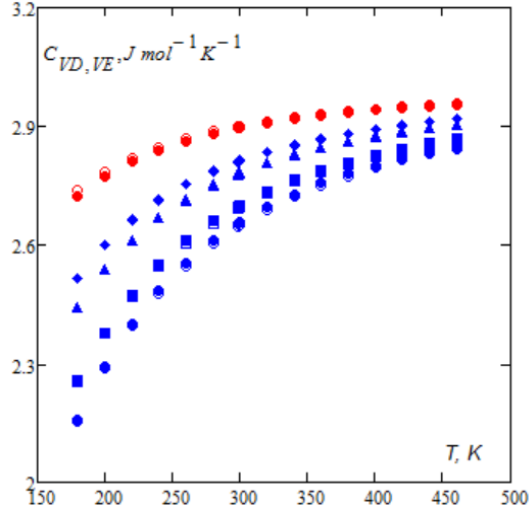

Fig. 3. The temperature dependence of the isochoric heat capacities $C_{VD}(v,T)$ (the open symbols) and $C_{VE}(v,T)$ (the filled symbols) (Eqs. 2 and 6) for $LiInS_2$ (the blue squares), $LiInSe_2$ (the blue diamonds), $LiGaS_2$ (the blue circles), $LiGaSe_2$ (the blue triangles) and $LiGaTe_2$ (the red circles).

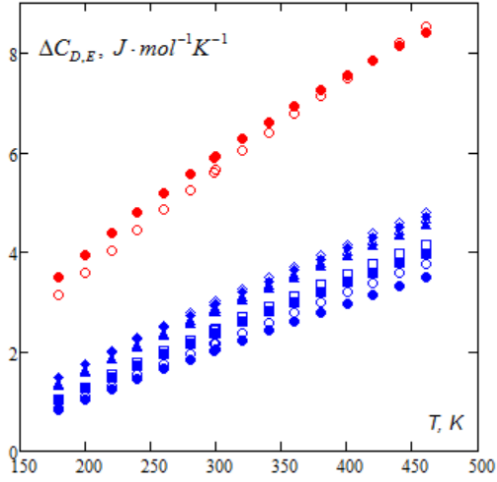

Fig. 4. The temperature dependence of the differences $\Delta C_D$ (the open symbols) and $\Delta C_E$ (the filled symbols) (Eqs. 4 and 8) for $LiInS_2$ (the blue squares), $LiInSe_2$ (the blue diamonds), $LiGaS_2$ (the blue circles), $LiGaSe_2$ (the blue triangles) and $LiGaTe_2$ (the red circles).

As one can see from Figs. 3 and 4 the models give the results for the isochoric heat capacity which are close to each other and they give the close results for the difference between the isobaric and isochoric heat capacities.

Fig. 5 shows the isochoric heat capacities $C_{VD}(v,T)$ and $C_{VE}(v,T)$ for $LiInS_2$, $LiInSe_2$, $LiGaS_2$, $LiGaSe_2$ and $LiGaTe_2$ which are calculated from Eqs. 2 and 6 using the values of the parameters $\theta_D$ and $\theta_E$ from Table 1 as a functions of the reduced temperatures $T/\theta_D$ and $T/\theta_E$, respectively.

The isobaric heat capacities $C_{PD}(v,T)$ and $C_{PE}(v,T)$ for $LiInS_2$, $LiInSe_2$, $LiGaS_2$, $LiGaSe_2$ and $LiGaTe_2$ which are calculated from Eqs. 3 and 7 using the values of the parameters $\theta_D$, $a_D$, $b_D$, $\theta_E$, $a_E$ and $b_E$ from Table 1 as the functions of the reduced temperatures $T/\theta_D$ and $T/\theta_E$, respectively, are shown on Fig. 6.

We can conclude from Fig. 5 that the isochoric heat capacities of $LiInS_2$, $LiInSe_2$, $LiGaS_2$, $LiGaSe_2$ and $LiGaTe_2$ (which has another structure than $LiInS_2$, $LiInSe_2$, $LiGaS_2$ and $LiGaSe_2$ [1]) as functions of the temperature reduced to the Debye (Einstein) temperature are described by Debye (Einstein) equation for the isochoric heat capacity.

Fig. 6 demonstrates that the isobaric heat capacities of the isostructural $LiInS_2$, $LiInSe_2$, $LiGaS_2$ and $LiGaSe_2$ as the functions of the temperature reduced to the Debay (Einstein) temperature are described by single Debay (Einstein) equation for the isobaric heat capacity.

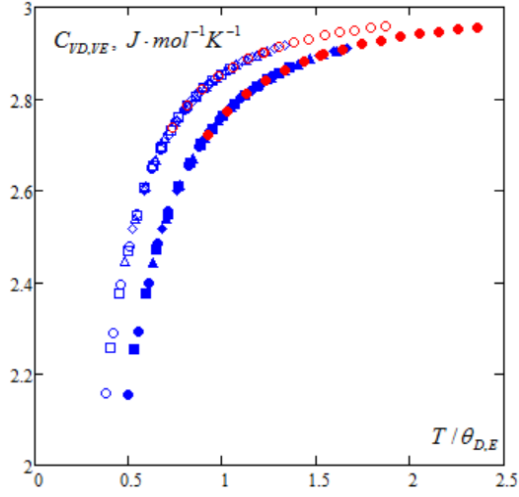

Fig. 5. The dependences of the isochoric heat capacities $C_{VD}$ on $T/\theta_D$ (the open symbols) and $C_{VE}$ on $T/\theta_E$ (the filled symbols) (Eqs. 2 and 6) on for $LiInS_2$ (the blue squares), $LiInSe_2$ (the blue diamonds), $LiGaS_2$ (the blue circles), $LiGaSe_2$ (the blue triangles) and $LiGaTe_2$ (the red circles).

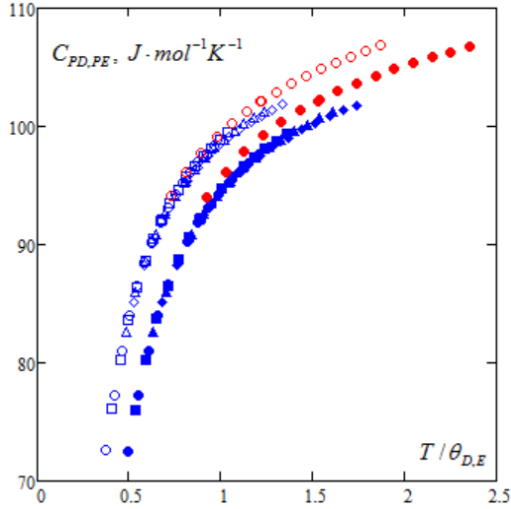

Fig. 6. The dependences of the isobaric heat capacities $C_{PD}$ on $T/\theta_D$ (the open symbols) and $C_{PE}$ on $T/\theta_E$ (the filled symbols) (Eqs. 3 and 7) for $LiInS_2$ (the blue squares), $LiInSe_2$ (the blue diamonds), $LiGaS_2$ (the blue circles), $LiGaSe_2$ (the blue triangles) and $LiGaTe_2$ (the red circles).

Fig. 7 presents the isochoric heat capacities $C_{VD}(v,T)$ and $C_{VE}(v,T)$ for $LiInS_2$, $LiInSe_2$, $LiGaS_2$, $LiGaSe_2$ and $LiGaTe_2$ which are calculated from Eqs. 2 and 6 using the values of the parameters $\theta_D$ and $\theta_E$ from Table 1 as the functions of the reduced temperatures $T/\theta_D$ (a), $T/\theta_E$ (b), $T/\lambda^{-1}\theta_D$ (c) and $T/\lambda\theta_E$ (d), respectively. Here $\lambda = \sqrt{5/3} \approx 1.29099$ is equal to the ratio $\omega_D/\omega_E$ if the values of $\omega_D$ and $\omega_E$ obey the condition

$$<\omega^2>_D = <\omega^2>_E \qquad (14)$$

of the equality of the means $<\omega^2>_D$ and $<\omega^2>_E$ of the squares $\omega^2$ of the frequency over the Debye and Einstein spectra, respectively. Indeed using the definitions

$$<\omega^2>_D \equiv \int_0^{\omega_D} \omega^2 g_D(v,\omega)d\omega / \int_0^{\omega_D} g_D(v,\omega)d\omega , \qquad (15)$$

$$<\omega^2>_E \equiv \int_0^\infty \omega^2 g_E(v,\omega)d\omega / \int_0^\infty g_E(v,\omega)d\omega \qquad (16)$$

and Eqs. 1, 5 and 14 one can easily show that $\omega_E/\omega_D = \sqrt{5/3}$.

As one can see from Fig. 7 the better single line for isochoric heat capacity corresponds to Figs. 7c and 7d, and the Debye and Einstein equations for the isochoric heat capacity of $LiInS_2$, $LiInSe_2$, $LiGaS_2$, $LiGaSe_2$ and $LiGaTe_2$ give the same results if the means of the squares of frequencies of the Debye and Einstein spectra are equal to each other.

The isobaric heat capacities $C_{PD}(v,T)$ and $C_{PE}(v,T)$ for $LiInS_2$, $LiInSe_2$, $LiGaS_2$, $LiGaSe_2$ and $LiGaTe_2$ which are calculated from Eqs. 2 and 6 using the values of the parameters $\theta_D$, $a_D$, $b_D$, $\theta_E$, $a_E$ and $b_E$ from Table 1 as the functions of the reduced temperatures $T/\theta_D$ (a), $T/\theta_E$ (b), $T/\lambda^{-1}\theta_D$ (c) and $T/\lambda\theta_E$ (d), respectively, are shown on Fig. 8.

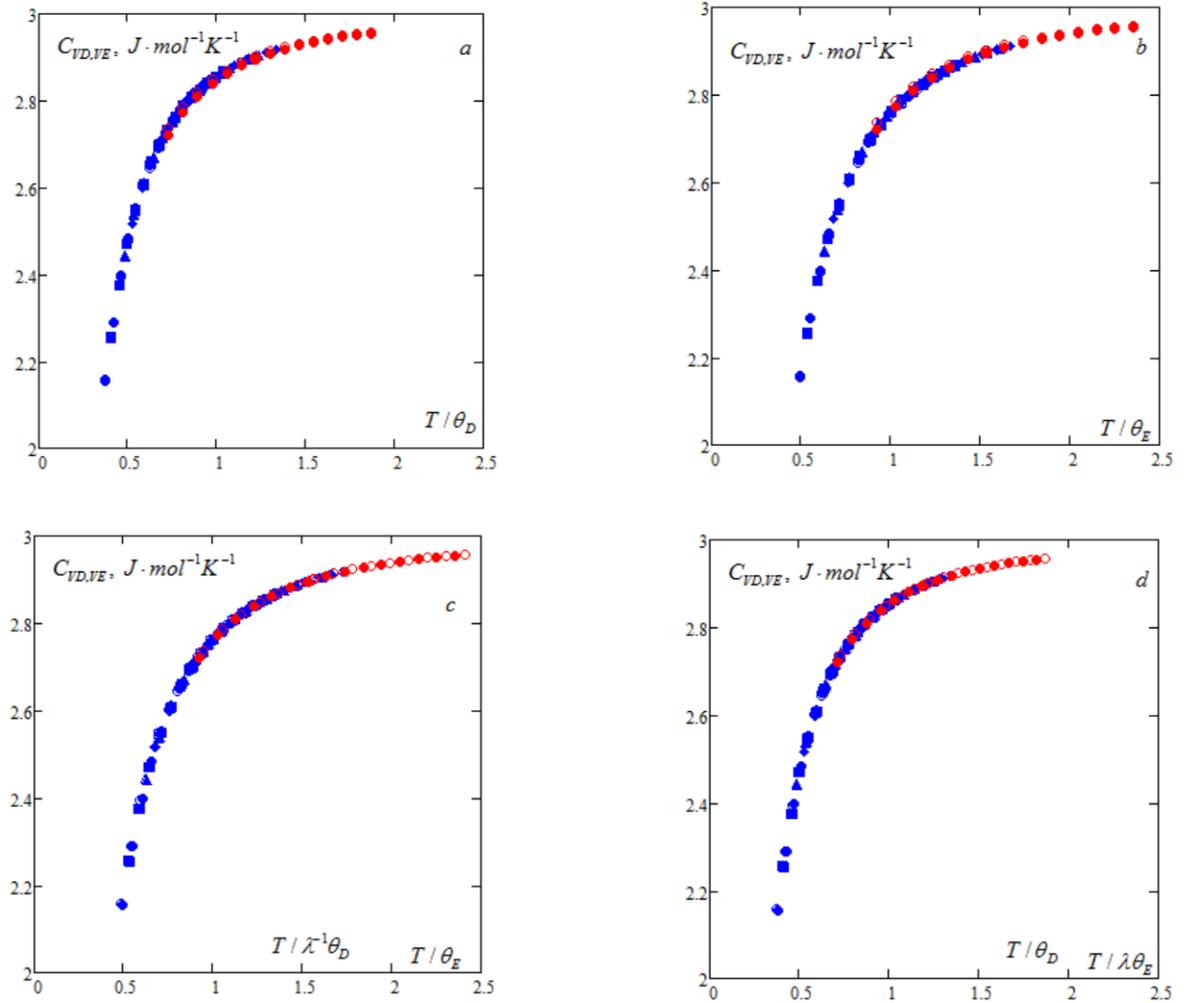

Fig. 7. The dependences of the isochoric heat capacities $C_{VD}$ (the open symbols) and $C_{VE}$ (the filled symbols) $T/\theta_D$ (a), $T/\theta_E$ (b), $T/\lambda^{-1}\theta_D$ (c) and $T/\lambda\theta_E$ (d), respectively, (Eqs. 3 and 7) for $LiInS_2$ (the blue squares), $LiInSe_2$ (the blue diamonds), $LiGaS_2$ (the blue circles), $LiGaSe_2$ (the blue triangles) and $LiGaTe_2$ (the red circles).

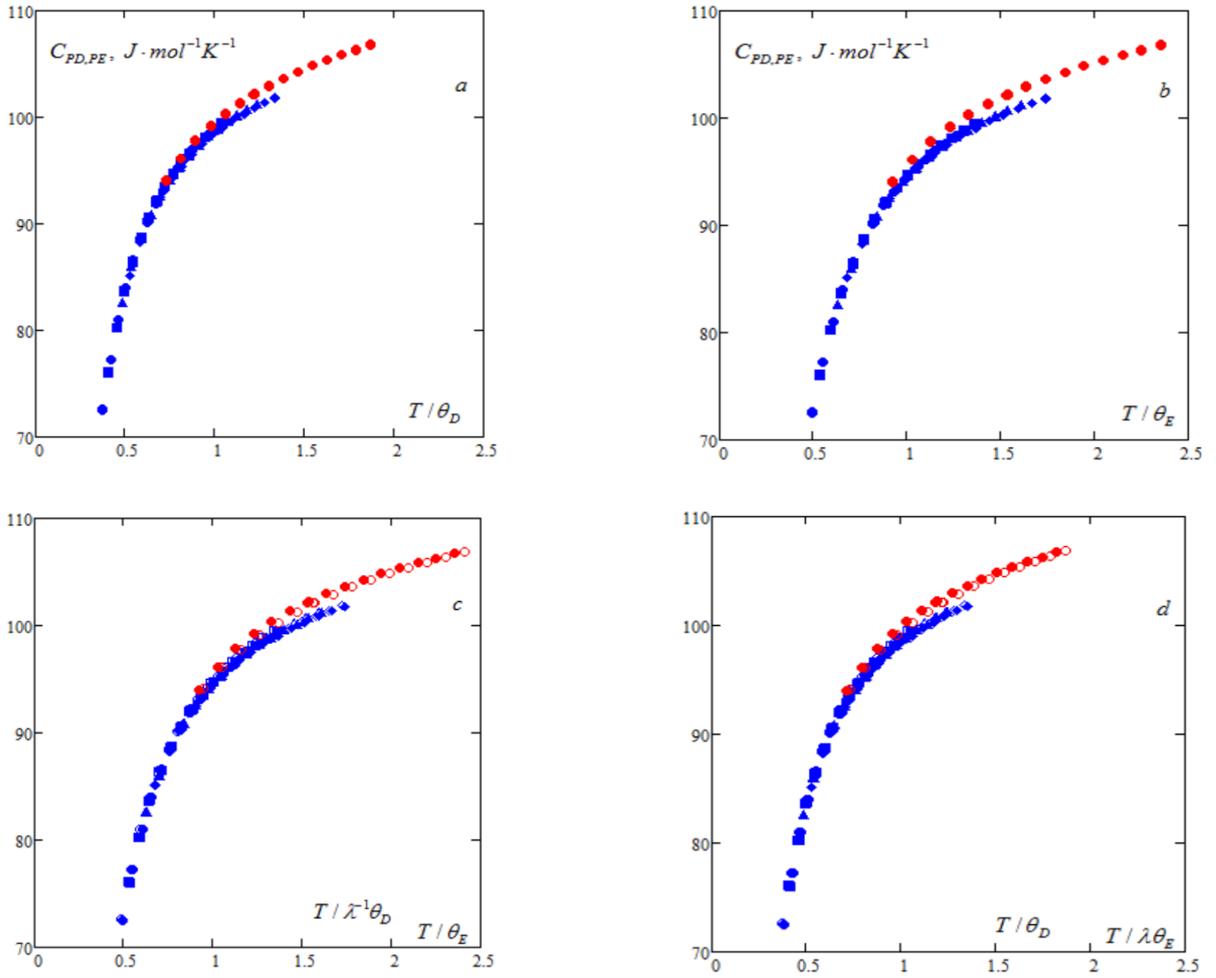

Fig. 8. The dependences of the isobaric heat capacities $C_{PD}$ (the open symbols) and $C_{PE}$ (the filled symbols) $T/\theta_D$ (a), $T/\theta_E$ (b), $T/\lambda^{-1}\theta_D$ (c) and $T/\lambda\theta_E$ (d), respectively, (Eqs. 3 and 7) for $LiInS_2$ (the blue squares), $LiInSe_2$ (the blue diamonds), $LiGaS_2$ (the blue circles), $LiGaSe_2$ (the blue triangles) and $LiGaTe_2$ (the red circles).

As one can see from Fig. 8 the better single line for isobaric heat capacity corresponds to Figs. 8a and 8b, and the Debye and Einstein equations for the isobaric heat capacity of $LiInS_2$, $LiInSe_2$, $LiGaS_2$ and $LiGaSe_2$ as the functions of the temperature reduced to the Debye or Einstein temperature give the same results.

The temperature dependences of the ratios $\Delta C_D/C_P$ and $\Delta C_E/C_P$ to the experimental isobaric heat capacity $C_P$ from [1], where the differences $\Delta C_D$ and $\Delta C_E$ between the isobaric and isochoric heat capacities are calculated using Eqs. 4 and 8 and the values of the parameters $\theta_D$, $a_D$, $b_D$, $\theta_E$, $a_E$ and $b_E$ from Table 1, are shown on Fig. 9.

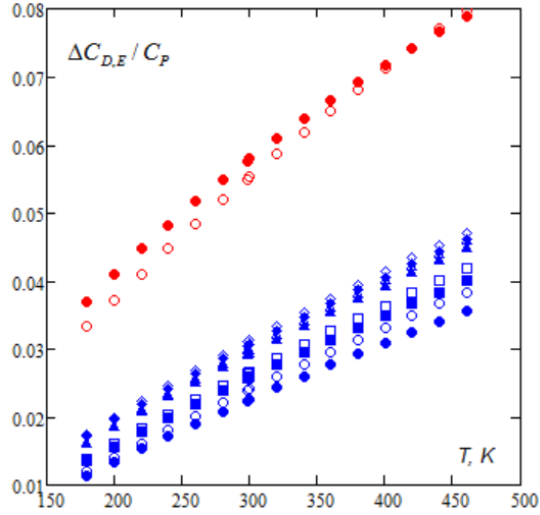

Fig. 9. The ratios $\Delta C_D / C_P$ (the open symbols) and $\Delta C_D / C_P$ (the filled symbols) of the isochoric heat capacities $C_{VD}$ and $C_{VE}$ (Eqs. 4 and 8) to the experimental data on $C_P$ [1] for $LiInS_2$ (the blue squares), $LiInSe_2$ (the blue diamonds), $LiGaS_2$ (the blue circles), $LiGaSe_2$ (the blue triangles) and $LiGaTe_2$ (the red circles).

One can see from Fig. 9 that the ratios increase with increasing temperature and

$0.01 < \Delta C_D / C_P < 0.05$, $0.01 < \Delta C_E / C_P < 0.05$

for $LiInS_2$, $LiInSe_2$, $LiGaS_2$ and $LiGaSe_2$,

$0.03 < \Delta C_D / C_P < 0.08$, $0.03 < \Delta C_E / C_P < 0.08$

for $LiGaTe_2$. Therefore we can conclude that it is necessary to take into account the contributions of $\Delta C_D$ and $\Delta C_E$ to the isobaric heat capacity because they are greater than the experimental uncertainties of [1].

**Table 2.** The values of the ratios $\theta_D / \theta_E$, $a_D / a_E$, $b_D / b_E$, $a_D / b_D$, $a_E / b_E$ and $\delta = (a_D + b_D)/(a_E + b_E)$.

| Compound | $\theta_D / \theta_E$ | $a_D / a_E$ | $b_D / b_E$ | $a_D / b_D$ | $a_E / b_E$ | $a_D + b_D$ | $a_E + b_E$ | $\delta$ |
|---|---|---|---|---|---|---|---|---|
| LiInS2 | 1.31 | 0.71 | 1.57 | 23.22 | 51.72 | 65.078 | 90.042 | 0.72 |
| LiInSe2 | 1.30 | 0.71 | 1.83 | 21.23 | 54.23 | 72.571 | 98.823 | 0.73 |
| LiGaS2 | 1.31 | 0.68 | 1.72 | 27.09 | 68.35 | 66.467 | 95.426 | 0.70 |
| LiGaSe2 | 1.30 | 0.72 | 1.84 | 24.97 | 63.73 | 70.182 | 95.021 | 0.74 |
| LiGaTe2 | 1.26 | 0.93 | 0.79 | 19.69 | 16.66 | 56.409 | 61.132 | 0.92 |

**Conclusions**

So we derived explicitly the equations for the isobaric heat capacity for the Debye and Einstein models of the phonon frequency spectrum in the present paper. We showed that: the data [1] for five chalcogenides $LiInS_2$, $LiInSe_2$, $LiGaS_2$, $LiGaSe_2$ and $LiGaTe_2$ are described by the Debye and Einstein models for the phonon frequency spectrum within their uncertainties; the models give the results for the isochoric heat capacity which are close to each other; the models give the close results for the difference between the isobaric and isochoric heat capacities; the isobaric heat capacities of the isostructural $LiInS_2$, $LiInSe_2$, $LiGaS_2$ and $LiGaSe_2$ as the functions of the temperature reduced to the Debye (Einstein) temperature are described by single Debay (Einstein) equation for the isobaric heat capacity; the isochoric heat capacities of $LiInS_2$, $LiInSe_2$, $LiGaS_2$, $LiGaSe_2$ and $LiGaTe_2$ (which has another structure than $LiInS_2$, $LiInSe_2$, $LiGaS_2$ and $LiGaSe_2$ [1]) as the functions of the temperature reduced to the

Debye (Einstein) temperature are described by the Debye (Einstein) equation for the isochoric heat capacity. It is shown also that the Debye and Einstein equations for the isochoric heat capacity of $LiInS_2$, $LiInSe_2$, $LiGaS_2$, $LiGaSe_2$ and $LiGaTe_2$ give the same results if the means of the squares of the frequencies of the Debye and Einstein spectra are equal to each other, and the Debye and Einstein equations for the isobaric heat capacity of $LiInS_2$, $LiInSe_2$, $LiGaS_2$ and $LiGaSe_2$ as the functions of the temperature reduced to the Debye or Einstein temperature give the same results.

As known the Einstein model of frequency spectrum do not describe the heat capacity of isotropic and anisotropic crystals at low temperatures [9-11]. Therefore we cannot extrapolate our results for the Einstein model to low temperature region.

The Debye model of the spectrum is valid for the isotropic system, such as polycrystalline materials [9-11,14]. The single crystal is anisotropic [14,15] and therefore it has the phonon frequency spectrum which differs from that of for isotropic system [14]. So the Debye spectrum is not valid for single crystals. Therefore we cannot extrapolate our results for the Debye spectrum to low temperatures.

The differences between results of [1] from that of [4,16] for $LiInSe_2$ and [4] for $LiInS_2$ were discussed in [1]. The differences may be related to the difference of their phonon spectra because the single crystals were investigated in [1] while the polycrystalline materials were studied in [4,16]. Therefore the compilation of the experimental data for single crystal [1] and polycrystalline LiInSe2 [16] as was done in [2] may be incorrect.

# Appendix

Let us consider the harmonic crystal consisting of $N$ atoms and having a volume $V$ with frequency spectrum $g(v,\omega)$, where $v = V/N$ is the volume per atom and $\omega$ is the frequency. The Helmholtz energy per atom $f(v,T)$ is defined by [9]

$$f(v,T) = u_0(v) + \frac{\hbar}{2}\int_0^\infty \omega g(v,\omega)d\omega + kT\int_0^\infty g(v,\omega)\ln\left[1-\exp\left(-\frac{\hbar\omega}{kT}\right)\right]d\omega, \quad (1)$$

where $T$ is the temperature, $u_0(v)$ is the internal energy per atom at $T=0$, $\hbar$ is the Planck constant, $k$ is Boltzmann constant. The frequency spectrum obeys the condition

$$\int_0^\infty \omega g(v,\omega)d\omega = 3. \quad (2)$$

We will use the following exact thermodynamic relations [9]

$$s(v,T) = -\left(\partial f(v,T)/\partial T\right)_v, \quad (3)$$
$$e(v,T) = f(v,T) + Ts(v,T), \quad (4)$$
$$C_V(v,T) = \left(\partial e(v,T)/\partial T\right)_v, \quad (5)$$
$$p(v,T) = -\left(\partial f(v,T)/\partial v\right)_T, \quad (6)$$
$$C_P(v,T) = C_V(v,T) - T\left(\partial p(v,T)/\partial T\right)_v^2 / \left(\partial p(v,T)/\partial v\right)_T, \quad (7)$$

where $p(v,T)$ is the pressure, and $s(v,T)$, $e(v,T)$, $C_V(v,T)$ and $C_P(v,T)$ are the entropy, internal energy, isochoric and isobaric heat capacities per atom, respectively.

**The Debye model**

The Debay spectrum $g_D(v,\omega)$ is defined by [9-11]

$$g_D(v,\omega) = 9\omega^2 \theta[\omega_D(v) - \omega]/\omega_D^3(v), \quad (8)$$

where $\omega_D(v)$ is the Debye frequency, $\theta(x)$ is the Heaviside step function: $\theta(x)=0$ if $x \leq 0$ and $\theta(x)=1$ if $x>0$. We have from Eqs. 1-8 eventually

$$f_D(v,T) = u_0(v) + \frac{9\hbar\omega_D(v)}{8} + \frac{9kT}{\omega_D^3(v)}\int_0^{\omega_D(v)} \omega^2 \ln\left[1-\exp\left(-\frac{\hbar\omega}{kT}\right)\right]d\omega, \quad (9)$$

$$s_D(v,T) = -\frac{9k}{\omega_D^3}\int_0^{\omega_D}\omega^2 \ln\left[1-\exp\left(-\frac{\hbar\omega}{kT}\right)\right]d\omega + \frac{9\hbar}{\omega_D^3 T}\int_0^{\omega_D}\omega^3\left[\exp\left(\frac{\hbar\omega}{kT}\right)-1\right]^{-1}d\omega, \quad (10)$$

$$e_D(v,T) = u_0(v) + \frac{9\hbar\omega_D(v)}{8} + \frac{9\hbar}{\omega_D^3(v)} \int_0^{\omega_D(v)} \omega^3 \left[\exp\left(\frac{\hbar\omega}{kT}\right) - 1\right]^{-1} d\omega, \tag{11}$$

$$C_{VD}(v,T) = 9k\left(\frac{T}{\theta_D}\right)^3 \cdot \int_0^{\theta_D/T} \frac{x^4 e^x}{(e^x - 1)^2} dx, \tag{12}$$

$$p_D(v,T) = -u_0' - \frac{9}{8}\hbar\omega_D' + \frac{27kT\omega_D'}{\omega_D^4} \int_0^{\omega_D} \omega^2 \ln\left[1 - \exp\left(-\frac{\hbar\omega}{kT}\right)\right] d\omega - \frac{9kT\omega_D'}{\omega_D} \ln\left[1 - \exp\left(-\frac{\hbar\omega_D}{kT}\right)\right],$$

$$p_D(v,T) = -u_0'(v) - \frac{9}{8}\hbar\omega_D'(v) - \frac{9\hbar\omega_D'(v)}{\omega_D^4(v)} \int_0^{\omega_D(v)} \omega^3 \left[\exp\left(\frac{\hbar\omega}{kT}\right) - 1\right]^{-1} d\omega, \tag{13}$$

$$\left(\frac{\partial p_D(v,T)}{\partial T}\right)_v = -\frac{9\hbar\omega_D'(v)}{\omega_D^4(v)} \int_0^{\omega_D(v)} \frac{\omega^4 \hbar}{kT^2} \exp\left(\frac{\hbar\omega}{kT}\right)\left[\exp\left(\frac{\hbar\omega}{kT}\right) - 1\right]^{-2} d\omega, \tag{14}$$

$$\left(\frac{\partial p_D(v,T)}{\partial T}\right)_v = -\frac{\theta_D'(v)}{\theta_D(v)} 9k\left(\frac{T}{\theta_D}\right)^3 \cdot \int_0^{\theta_D/T} \frac{x^4 e^x}{(e^x - 1)^2} dx, \tag{15}$$

$$\left(\frac{\partial p_D(v,T)}{\partial T}\right)_v = -\frac{\theta_D'(v)}{\theta_D(v)} C_V(v,T) = -\frac{\omega_D'(v)}{\omega_D(v)} C_V(v,T), \tag{16}$$

$$\left(\frac{\partial p_D(v,T)}{\partial v}\right)_T = -u_0'' - \frac{9}{8}\hbar\omega_D'' - \frac{9\hbar\omega_D''}{\omega_D^4} \int_0^{\omega_D} \omega^3 \left[\exp\left(\frac{\hbar\omega}{kT}\right) - 1\right]^{-1} d\omega +$$
$$\frac{36\hbar(\omega_D')^2}{\omega_D^5} \int_0^{\omega_D} \omega^3 \left[\exp\left(\frac{\hbar\omega}{kT}\right) - 1\right]^{-1} d\omega - \frac{9\hbar(\omega_D')^2}{\omega_D} \left[\exp\left(\frac{\hbar\omega_D}{kT}\right) - 1\right]^{-1}, \tag{17}$$

$$\left(\frac{\partial p_D(v,T)}{\partial v}\right)_T = -u_0'' - \frac{9}{8}\hbar\omega_D'' - \frac{9\hbar\omega_D''}{\omega_D^4} \int_0^{\omega_D} \omega^3 \left[\exp\left(\frac{\hbar\omega}{kT}\right) - 1\right]^{-1} d\omega +$$
$$T\frac{(\omega_D')^2}{\omega_D^2} 9k\left(\frac{T}{\theta_D}\right)^3 \cdot \int_0^{\theta_D/T} \frac{x^4 e^x}{(e^x - 1)^2} dx, \tag{18}$$

$$\left(\frac{\partial p_D(v,T)}{\partial v}\right)_T = -u_0'' - \frac{9}{8}\hbar\omega_D'' - \frac{9\hbar\omega_D''}{\omega_D^4} \int_0^{\omega_D} \omega^3 \left[\exp\left(\frac{\hbar\omega}{kT}\right) - 1\right]^{-1} d\omega + T\frac{(\omega_D')^2}{\omega_D^2} C_V(v,T), \tag{19}$$

$$\left(\frac{\partial p_D(v,T)}{\partial v}\right)_T = -u_0'' - \frac{9}{8}\hbar\omega_D'' - \frac{9\hbar\omega_D''}{4}\left[\exp\left(\frac{\hbar\omega_D}{kT}\right) - 1\right]^{-1} -$$
$$\frac{9\hbar\omega_D''}{4\omega_D^4} \int_0^{\omega_D} \omega^4 \frac{\hbar}{kT} \exp\left(\frac{\hbar\omega}{kT}\right)\left[\exp\left(\frac{\hbar\omega}{kT}\right) - 1\right]^{-2} d\omega + T\frac{(\omega_D')^2}{\omega_D^2} C_V(v,T), \tag{20}$$

$$\left(\frac{\partial p_D(v,T)}{\partial v}\right)_T = -u_0'' - \frac{9}{8}\hbar\omega_D'' - \frac{9\hbar\omega_D''}{4}\left[\exp\left(\frac{\hbar\omega_D}{kT}\right) - 1\right]^{-1} -$$
$$\frac{T\omega_D''}{4\omega_D} 9k\left(\frac{T}{\theta_D}\right)^3 \cdot \int_0^{\theta_D/T} \frac{x^4 e^x}{(e^x - 1)^2} dx + T\frac{(\omega_D')^2}{\omega_D^2} C_V(v,T), \tag{21}$$

$$\left(\frac{\partial p_D(v,T)}{\partial v}\right)_T = -u_0'' - \frac{9}{8}\hbar\omega_D'' - \frac{9\hbar\omega_D''}{4}\left[\exp\left(\frac{\hbar\omega_D}{kT}\right) - 1\right]^{-1} + \left(\frac{(\omega_D')^2}{\omega_D^2} - \frac{\omega_D''}{4\omega_D}\right) T C_{VD}(v,T), \tag{22}$$

$$C_{PD}(v,T) = C_{VD} + \frac{\left(\frac{\omega'_D}{\omega_D}\right)^2 TC_{VD}^2(v,T)}{u''_0 + \frac{9}{8}\hbar\omega''_D + \frac{9\hbar\omega''_D}{4}\left[\exp\left(\frac{\hbar\omega_D}{kT}\right)-1\right]^{-1} - \left(1-\frac{\omega''_D\omega_D}{4(\omega'_D)^2}\right)\frac{(\omega'_D)^2}{\omega_D^2}TC_{VD}}, \quad (23)$$

$$C_{PD} = C_{VD}\left(1 + \frac{\frac{\hbar(\omega'_D)^2}{\omega_D}\frac{kT}{\hbar\omega_D}\frac{C_{VD}(v,T)}{k}}{u''_0 + \frac{9}{8}\hbar\omega''_D + \frac{9\hbar\omega''_D}{4}\left[\exp\left(\frac{\hbar\omega_D}{kT}\right)-1\right]^{-1} - \left(1-\frac{\omega''_D\omega_D}{4(\omega'_D)^2}\right)\frac{\hbar(\omega'_D)^2}{\omega_D}\frac{kT}{\hbar\omega_D}\frac{C_{VD}}{k}}\right), \quad (24)$$

$$C_{PD}(v,T) = C_{VD}(v,T) + \frac{(T/\theta_D)\cdot[C_{VD}^2(v,T)/k]}{a_D + \frac{9b_D}{8} + \frac{9b_D}{4}\left[\exp\left(\frac{\theta_D}{T}\right)-1\right]^{-1} - \frac{T}{\theta_D}\left(1-\frac{b_D}{4}\right)\cdot\frac{C_{VD}(v,T)}{k}}, \quad (25)$$

where $\theta_D = \theta_D(v) = \hbar\omega_D(v)/k$ is the Debye temperature, $\theta'_D = \theta'_D(v) = \hbar\omega'_D(v)/k$, $\theta''_D = \hbar\omega''_D(v)/k$, $u'_0 = u'_0(v) = du_0/dv$, $u''_0 = u''_0(v) = d^2u_0(v)/dv^2$, $\omega'_D = \omega'_D(v) = d\omega_D(v)/dv$, $\omega''_D = \omega''_D(v) = d^2\omega_D(v)/dv^2$, $a_D = a_D(v) = \omega_D u''_0/\hbar(\omega'_D)^2$, $b_D = b_D(v) = \omega_D \omega''_D/(\omega'_D)^2$.

We have for the difference between the isobaric and isochoric heat capacities $\Delta C_D(v,T) = C_{PD}(v,T) - C_{VD}(v,T)$ the relation

$$\Delta C_D(v,T) = \frac{(T/\theta_D)\cdot[C_{VD}^2(v,T)/k]}{a_D + \frac{9b_D}{8} + \frac{9b_D}{4}\left[\exp\left(\frac{\theta_D}{T}\right)-1\right]^{-1} - \frac{T}{\theta_D}\left(1-\frac{b_D}{4}\right)\cdot\frac{C_{VD}(v,T)}{k}}. \quad (26)$$

The relation

$$C_{VD}(v,T) = k\frac{12\pi^4}{5}\left(\frac{T}{\theta_D}\right)^3 \quad (27)$$

is valid for $T \ll \theta_D$ [9-11] therefore one can obtain from Eq. 31 the relation

$$C_{PD}(v,T) = k\frac{12\pi^4}{5}\left(\frac{T}{\theta_D}\right)^3\left(1 + \frac{12\pi^4/5}{a_D + 9b_D/8}\left(\frac{T}{\theta_D}\right)^4\right) \quad (28)$$

which is valid at low temperatures.

**Einstein model**

The Einstein frequency spectrum is defined by [10,12]

$$g_E(v,\omega) = 3\delta[\omega - \omega_E(v)], \quad (29)$$

where $\omega_E(v)$ is the Einstein frequency. We have from Eqs. 1-7 and 29

$$f_E(v,T) = u_0(v) + \frac{3\hbar\omega_E(v)}{2} + 3kT\ln\left[1-\exp\left(-\frac{\hbar\omega_E(v)}{kT}\right)\right], \quad (30)$$

$$s_E(v,T) = -3k\ln\left[1-\exp\left(-\frac{\hbar\omega_E(v)}{kT}\right)\right] + 3\frac{\hbar\omega_E(v)}{T}\left[\exp\left(\frac{\hbar\omega_E(v)}{kT}\right)-1\right]^{-1}, \quad (31)$$

$$e_E(v,T) = u_0(v) + \frac{3\hbar\omega_E(v)}{2} + 3\hbar\omega_E(v)\left[\exp\left(\frac{\hbar\omega_E(v)}{kT}\right) - 1\right]^{-1}, \tag{32}$$

$$C_{VE}(v,T) = 3k\left(\frac{\theta_E(v)}{T}\right)^2 \exp\left(\frac{\theta_E(v)}{T}\right)\left[\exp\left(\frac{\theta_E(v)}{T}\right) - 1\right]^{-2}, \tag{33}$$

$$p_E(v,T) = -u_0'(v) - \frac{3\hbar\omega_E'(v)}{2} - 3\hbar\omega_E'(v)\left[\exp\left(\frac{\hbar\omega_E(v)}{kT}\right) - 1\right]^{-1}, \tag{34}$$

$$\left(\frac{\partial p_E(v,T)}{\partial T}\right)_v = -\frac{\omega_E'(v)}{\omega_E(v)} C_{VE}(v,T), \tag{35}$$

$$\left(\frac{\partial p_E(v,T)}{\partial v}\right)_T = -u_0''(v) - \frac{3\hbar\omega_E''(v)}{2} - 3\hbar\omega_E''(v)\left[\exp\left(\frac{\hbar\omega_E(v)}{kT}\right) - 1\right]^{-1} +$$
$$\frac{3\hbar^2[\omega_E'(v)]^2}{kT}\exp\left(\frac{\hbar\omega_E(v)}{kT}\right)\left[\exp\left(\frac{\hbar\omega_E(v)}{kT}\right) - 1\right]^{-2}, \tag{36}$$

$$\left(\frac{\partial p_E(v,T)}{\partial v}\right)_T = -u_0''(v) - \frac{3\hbar\omega_E''(v)}{2} - 3\hbar\omega_E''(v)\left[\exp\left(\frac{\hbar\omega_E(v)}{kT}\right) - 1\right]^{-1} + \frac{\hbar[\omega_E'(v)]^2}{\omega_E(v)} \frac{T}{\theta_E(v)} \frac{C_V(v,T)}{k}, \tag{37}$$

$$C_{PE}(v,T) = C_{VE}(v,T) + \frac{[T/\theta_E(v)] \cdot [C_{VE}^2(v,T)/k]}{a_E(v) + \frac{3b_E(v)}{2} + 3b_E(v)\left[\exp\left(\frac{\theta_E(v)}{T}\right) - 1\right]^{-1} - \frac{T}{\theta_E(v)} \cdot \frac{C_{VE}(v,T)}{k}}, \tag{38}$$

where $\theta_E = \theta_E(v) = \hbar\omega_E(v)/k$ is the Einstein temperature, $\theta_E' = \hbar\omega_E'(v)/k$, $\theta_E'' = \hbar\omega_E''(v)/k$, $\omega_E' = \omega_E'(v) = d\omega_E(v)/dv$, $\omega_E'' = \omega_E''(v) = d^2\omega_E(v)/dv^2$, $a_E = a_E(v) = \omega_E u_0''/\hbar(\omega_E')^2$ and $b_E = b_E(v) = \omega_E \omega_E''/(\omega_E')^2$.

It is easy to see from Eqs. 41 and 46 that the relations are valid

$$C_{VE}(v,T) = 3k\left(\frac{\theta_E}{T}\right)^2 \exp\left(-\frac{\theta_E}{T}\right)\left[1 + 2\exp\left(-\frac{\theta_E}{T}\right)\right], \tag{39}$$

$$C_{PE}(v,T) = 3k\left(\frac{\theta_E}{T}\right)^2 \exp\left(-\frac{\theta_E}{T}\right)\left[1 + \left(\frac{3\theta_E/T}{a_E + 3b_E/2} + 2\right)\exp\left(-\frac{\theta_E}{T}\right)\right] \tag{40}$$

are valid for $T \ll \theta_E$.

If $\omega_E(v) = 0$ then one can conclude from Eq. 32 that

$$e_E(v,T) = u_0(v) + 3kT. \tag{41}$$

One can obtain from Eq. 32 the relation

$$T(e,v) = \frac{\hbar\omega_E(v)/k}{\ln\left(\frac{e - u_0(v) + 3\hbar\omega_E(v)/2}{e - u_0(v) - 3\hbar\omega_E(v)/2}\right)}, \tag{42}$$

to define temperature as a function of internal energy and volume for $\omega_E(v) > 0$ and

$$e > u_0(v) + 3\hbar\omega_E(v)/2. \tag{43}$$

We conclude from Eq. 42 that $T = 0$, if

$$e = u_0(v) - 3\hbar\omega_E(v)/2. \tag{44}$$

If $u_0(v) + 3\hbar\omega_E(v)/2 < 0$ then it is easy to see from Eq. 58 that for $e = 0$

$$T(e=0,v) = T_e(v) \equiv \frac{\hbar\omega_E(v)/k}{\ln\left(\dfrac{-u_0(v) + 3\hbar\omega_E(v)/2}{-u_0(v) - 3\hbar\omega_E(v)/2}\right)}. \tag{45}$$

We conclude from Eq. 45 that $e = 0$ and $T = 0$ if

$$u_0(v) = 3\hbar\omega_E(v)/2. \tag{46}$$

We conclude from Eqs. 34-35 that

$$p_E(v,T) = -u_0'(v), \tag{47}$$

$$\left(\frac{\partial p_E(v,T)}{\partial T}\right)_v = 0, \tag{48}$$

if $\omega_E'(v) = 0$, $T > 0$ and $\omega_E(v) > 0$.

If $\omega_E'(v) \neq 0$, $\omega_E(v) > 0$ and

$$\frac{-3\hbar\omega_E'(v)}{p + u_0'(v) + 3\hbar\omega_E'(v)/2} > 0, \tag{49}$$

then we have from Eq. 34 the relation

$$T(p,v) = \frac{\hbar\omega_E(v)/k}{\ln\left(\dfrac{p + u_0'(v) - 3\hbar\omega_E'(v)/2}{p + u_0'(v) + 3\hbar\omega_E'(v)/2}\right)}. \tag{50}$$

to define temperature as the function of pressure and volume. We conclude from Eq. 49 that there are two cases

$$\omega_E'(v) < 0, \ p + u_0'(v) + 3\hbar\omega_E'(v)/2 > 0, \tag{51}$$

$$\omega_E'(v) > 0, \ p + u_0'(v) + 3\hbar\omega_E'(v)/2 < 0. \tag{52}$$

We conclude from Eq. 50 that $T = 0$ if

$$p = -u_0'(v) + 3\hbar\omega_E'(v)/2. \tag{53}$$

We have from Eq. 50

$$T(p=0,v) = T_0(v) \equiv \frac{\hbar\omega_E(v)/k}{\ln\left(\dfrac{u_0'(v) - 3\hbar\omega_E'(v)/2}{u_0'(v) + 3\hbar\omega_E'(v)/2}\right)}. \tag{54}$$

We can conclude from Eq. 54 that $p = 0$ and $T = 0$ if

$$u_0'(v) = 3\hbar\omega_E'(v)/2. \tag{55}$$

We can conclude from Eq. 54 that

$$T(p,v) = p\frac{\omega_E(v)}{3|\omega'_E(v)|k} \tag{56}$$

for $\omega'_E(v) < 0$ and $p >> |u'_0(v) + 3\hbar\omega'_E(v)/2|$ or $\omega'_E(v) > 0$ and $p >> |u'_0(v) - 3\hbar\omega'_E(v)/2|$.

We have from Eq. 34 for $T << \theta_E$

$$p_E(v,T) = -u'_0(v) - \frac{3\hbar\omega'_E(v)}{2} - 3\hbar\omega'_E(v)\exp\left(-\frac{\hbar\omega_E(v)}{kT}\right). \tag{57}$$